\documentclass[oldversion]{aa}
\usepackage{txfonts}
\usepackage{natbib, graphicx}

\begin{document}
\title{Parity Dependence in Strong Lens Systems as a Probe of Dark Matter Substructure}
\author{Jacqueline Chen}
\institute{Argelander-Institut f\"{u}r Astronomie, Universit\"{a}t Bonn,
Auf dem H\"{u}gel 71,
D-53121 Bonn; 
       {\tt jchen@astro.uni-bonn.de}
} 
\titlerunning{Parity Dependence}

\abstract{The amount of mass in small, dark matter clumps within galaxies (substructure) is an important test of cold dark matter.  One approach to measuring the substructure mass fraction is to analyze the fluxes of images that have been strongly lensed by a galaxy.  Flux ratios between images that are anomalous with respect to smooth (no substructure) models have previously suggested that there is a greater amount of substructure than found in dark matter simulations.  One measure of anomalous flux ratios is parity dependence -- that the fluxes of different images of a source are perturbed differently.  In this paper, we discuss parity dependence as a probe of dark matter substructure.  We find that reproducing the observed parity dependence requires a significant alignment between concentrated dark matter clumps and images.  The results may imply a larger fraction of mass in substructures than suggested by some dark matter simulations and that the observed parity dependence is unlikely to be reproduced by luminous satellites of lens galaxies.  
}
\maketitle

\section{Introduction}

Strong gravitational lenses are important probes of the mass distribution in the universe and have been used to measure both cosmological parameters and to understand and constrain the structure of dark matter halos and subhalos.  One particular problem in cold dark matter (CDM) that may be addressed with strong gravitational lensing is the discrepancy between the relatively small number of observed satellite galaxies around galaxies such as our own and the numerous small-mass clumps seen in simulations of dark matter halos \citep{klypin_etal99,moore_etal99,gao_etal04}.  \citet{dalal_kochanek02} attempted to constrain the fraction of mass in dark matter substructures using radio observations of quasar sources that have been quadruply-lensed by galaxies.  In such systems, dark matter clumps may perturb the magnification of images, inducing both flux ratios among the images that are not well-fit by smooth lens models and small astrometric perturbations to the image positions \citep{mao_schneider98,metcalf_madau01}.  Radio observations are necessary in order to avoid contamination from perturbations from stars in the lensing galaxies.   \citet{dalal_kochanek02} found a surface mass fraction between 0.6\% and 7\% within 90 percent confidence intervals.  This result is marginally consistent with the average surface mass density in substructures at the projected separation of the images as estimated by dark matter simulations at $z=0$ of $f_{\rm sub}=0.5\%$ \citep{mao_etal04}.   

An alternate method to investigating the substructure mass fraction compares observations and simulations directly, bypassing lens modeling.  For lens systems with four images of the source, a lens configuration where three of the images are found near one another is referred to as a cusp.  For a smooth lens halo, the cusp relation suggests that the sum of the fluxes in the outer two nearby images should be similar to that of the central image.  Several papers compare deviations from the cusp relation in observations and simulations and suggest that there is more substructure in the observations than can be accounted for in dark matter simulations \citep{maccio_etal06,amara_etal06}.

In anomalous flux ratio lens systems, the distribution of magnification perturbations is not the same for each image of a four-image lens:  the amount by which the image is brightened or dimmed relative to the best-fit lens model is dependent on the parity of the image.  Image parity describes the relative orientation of an image to the source, and -- as it is equivalent to the signs of the eigenvalues of the magnification matrix -- all images can be described as having either positive parity (double-positive or double-negative eigenvalues) or negative parity.  Parity dependence in radio observations has been used previously as evidence that anomalous flux ratios are due to dark matter substructure and not to any propagation effect, whose effects would be parity independent in general.    \citet{kochanek_dalal04} find that the parity dependence in the flux anomalies is such that {\it the brightest negative parity image is preferentially made dimmer by substructures} when compared to the best-fit smooth macromodel.  

On the theoretical side, \citet{keeton01} makes simple estimates of the effects of perturbers, ignoring the effects of perturbations to the image positions and using a linear analysis for the magnification perturbations.  He finds that for singular isothermal sphere (SIS) perturbers, double-positive parity images may only be brightened while double-negative parity images may only be dimmed and negative parity images may both be brightened or dimmed.  It is possible that this may result in an averaged behavior similar to that observed by \citet{kochanek_dalal04}.  The parity dependence described by \citet{keeton01}, however, is limited to untruncated SISs.  \citet{rozo_etal06}, using a similar analysis, calculate the magnification perturbations for more realistic clumps and study their behavior in statistical samples.  They find that for a substructure mass fraction of 0.5\%, the average magnification perturbation is small ($\lesssim 1\%$), and it is very slightly more likely that negative parity images are made brighter by substructure and positive parity images are made dimmer, an averaged result in contrast to the observations.  

The difficulties in reproducing the observed parity dependence suggest that these simple linear perturbation estimates are insufficient, and astrometric perturbations are significant and/or substructure causes non-linear perturbations, both of which would require lens modeling to investigate\footnote{While the estimates of the size of astrometric perturbations are relatively small (milliarcsecond scale) \citep{chen_etal07,metcalf_madau01}, they may affect the best-fit macromodel and the magnification perturbations in significant ways.}.  In this paper, we attempt to reproduce the observed parity dependence by modeling mock lens systems with various distributions of substructure.  In addition to resolving the contradictions between observations and theoretical estimates, we attempt to constrain the kinds of substructure models that will reproduce the observed parity dependence.  We show that the distribution of substructure necessary is inconsistent both with the results of \citet{mao_etal04} and with the anomalous flux ratios being solely due to observed luminous satellite galaxies.  

The paper is organized as follows.  Section \ref{sec:lens_model} describes the lens modeling, while Section \ref{sec:observations} discusses the observational sample and reproduces the previously observed parity dependence.  Section \ref{sec:mock} describes the substructure models used and how mock observations are created.  Section \ref{sec:results} presents the results of testing several substructure models and compares them to the observational sample.  Discussion and conclusions are presented in Sections \ref{sec:discussion} and \ref{sec:conclusions}.  

\section{Lens Modeling}
\label{sec:lens_model}

We model the positions of lensed images using an automated lens fitting code with a singular isothermal ellipsoid (SIE) mass distribution and an external shear component.  The 9 model parameters are 1.) the Einstein radius $b$;  2.) the projected axis ratio $q$;  3.) the orientation of the halo ellipticity $\theta_{q}$;  4.) the external shear $\gamma$ which describes the effect of structure near the lens halo (e.g., a nearby group of galaxies);  5.) the orientation of the shear $\theta_{\gamma}$;  6.) \& 7.) the source position, $x_{\rm source}$ and $y_{\rm source}$;  and 8.) \& 9.) the center of lens, $x_{\rm center}$ and $y_{\rm center}$.  The SIE has a projected surface density
\begin{equation}
\kappa(\xi) = \frac{b}{2\xi},
\end{equation}
where $\xi$ is the elliptical coordinate satisfying $\xi^2 = x^2+ y^2/q^2$ and where $x$ and $y$ are the Cartesian coordinates and $q$ is the axis ratio.  In the limit of circular symmetry, $b$, the Einstein radius, is related to the 1-d velocity dispersion $\sigma$ by
\begin{equation}
b=4\pi\left(\frac{\sigma}{c}\right)^2 \frac{D_{\rm ls}}{D_{\rm os}},
\end{equation}
where $c$ is the speed of light in a vacuum, and $D_{\rm ls}$ and $D_{\rm os}$ are the angular diameter distances from the lens to the source and from the observer to the source, respectively.

We restrict ourselves to lensed systems with four observed images.  The observed constraints consist of 8 coordinates of the images and a constraint from the observed center of the lens potential (presumably the position of the lensing galaxy), $x_{\rm center}$ and $y_{\rm center}$, for a total of 10 constraints.  Fluxes are not included as constraints.

The lens modeling algorithm requires several steps and is based upon the publicly available {\it gravlens} lens modeling code \citep{keeton01b}.  We employ a downhill simplex minimization routine,  and we first find best-fit parameters first in the source plane, which is faster than fitting in the image plane.\footnote{We fit in the source plane by minimizing the dispersion in the modeled source positions.  Further details are found in Appendix \ref{sec:appendix}.}  We start by finding appropriate values for the lens halo parameters -- $b$, $q$, and $\theta_{q}$ -- then fixing the lens halo parameters and finding appropriate values for the external shear $\gamma$ and $\theta_{\gamma}$.  We do several iterations of fits for all the parameters in the source plane.  At each iteration, we slightly perturb the image positions.  Perturbing the image positions gives us several sets of data with formally identical observational constraints and may help to avoid ending in local minima.  Finally, we do several iterations of fitting all the parameters in the image plane.  Additional details are found in Appendix \ref{sec:appendix}.

\section{Observed Lens Systems}
\label{sec:observations}

We use a set of observed lens systems both to test the observed parity dependence and as the basis for mock observations.  \citet{dalal_kochanek02} use 7 four-image lens systems with anomalous flux ratios;  their sample also represents a fair fraction of all 4-image systems with galaxy lenses observed in the radio.  \citet{kochanek_dalal04} use those 7 systems with the exception of PG1115+080 and add two more systems in order to test the parity dependence of anomalous flux ratio systems.  In this paper, we use 5 of the 7 lens systems used in \citet{dalal_kochanek02}, excluding PG1115+080 which does not have observed radio fluxes and B1608+656 which is a  double lens system.  The systems used are  MG0414+0534 \citep{katz_etal97,ros_etal00}, B0712+472 \citep{jackson_etal98}, B1422+231\citep{patnaik_etal99,patnaik_etal92}, B1933+503 \citep{sykes_etal98}, and B2045+265 \citep{mckean_etal07}.\footnote{Additional data taken from http://cfa-www.harvard.edu/castles.}  

\begin{figure}[h]
\centering
\resizebox{3.in}{!}
	{\includegraphics{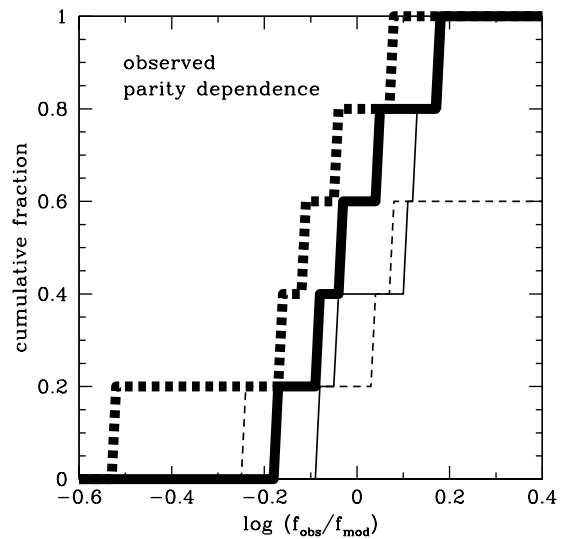}}
\caption{The cumulative flux perturbation distribution for 5 observed lens systems.  Solid lines represent positive parity images, while dashed represent negative parity images.  The brighter of two  images with the same parity is denoted by a thick line and the fainter by a thin line. \label{fig:data}}
\end{figure}

Using our automated lens modeling code, we try to reproduce the results of \citet{kochanek_dalal04}.  As discussed previously, we fit the positions of the images and the lensing galaxy (10 constraints) with a SIE lens plus external shear.  The fluxes (and magnifications) of the images are not used as constraints.  The estimated flux perturbation for each image is defined by
\begin{equation}
\delta_{f,i} = {\rm log} (f_{{\rm obs,} i}/f_{{\rm mod,} i}),
\end{equation}
where $f_{{\rm obs,} i}$ is the observed flux of the $i$th image and $f_{{\rm mod,} i}$ is the modeled flux of the $i$th image.  The modeled fluxes are related to the modeled magnifications by
\begin{equation}
f_{{\rm mod,} i} = S_{0} ~\mu_{{\rm mod,}i}
\end{equation}
where $S_0$ is the unlensed source flux.  The unlensed source flux, however, is unknown.  If there were no flux perturbations, the source flux would be equivalent to $S_0 = f_{{\rm obs,} i}/\mu_{{\rm mod,}i}$.  It can be estimated by 
\begin{equation}
\tilde{S}_0 = \frac{1}{2} \sum_{i=1}^2 \frac{f_{{\rm obs,} i}}{\mu_{{\rm mod,}i}},
\end{equation}
where -- in order to minimize the biasing due to flux perturbations -- we discard the images with the largest and smallest ratios of $f_{{\rm obs,} i}/\mu_{{\rm mod,}i}$  and average over the remaining two images.\footnote{\citet{kochanek_dalal04} average over all four images.}

We separate the four images of each lens by brightness and parity, denoting positive and negative parity images by ``$+$" and ``$-$" signs and brightest and faintest images by ``$\bigstar$" and ``$\ast$" signs.  Thus, each lens has a brightest positive parity image ($\mathcal I_{+\bigstar}$), a faintest positive parity image ($\mathcal I_{+\ast}$), a brightest negative parity image ($\mathcal I_{-\bigstar}$), and a faintest negative parity image ($\mathcal I_{-\ast}$).  Correspondingly, each image, $\mathcal I_i$, has a magnification, $\mu_i$, and flux perturbation, $\delta_{f,i}$, or magnification perturbation, $\delta_i= {\rm log} (\mu_{{\rm obs,}i}/\mu_{{\rm mod,}i})$.

The cumulative distribution of flux perturbations, $N(< \delta_{f,i})/(\rm total~number~of~lenses)$, is shown in Figure \ref{fig:data}.  The results are plotted so that negative (positive) values represent images with fluxes that are dimmer (brighter) than the best-fit modeled fluxes.  The results do not reproduce the exact relation seen in \citet{kochanek_dalal04}:  specifically, our results show overall larger magnification perturbations in the faint images (in particular in $\mathcal I_{-\ast}$).  In addition, not all of our lenses show that $\delta_{f,-\bigstar} <  0$ as is the case in \citet{kochanek_dalal04}.  This does not constitute a discrepancy since the unlensed source flux has been estimated slightly differently.  Our results generally agree with \citet{kochanek_dalal04} as parity dependence is very clearly seen and is such that  $\delta_{f,-\bigstar} <  \delta_{f, \rm ~all~ others}$ (i.e., the brightest negative parity image is less magnified or more demagnified relative to its modeled magnification than other images are compared to their modeled magnifications).  Our code uses commonly used methods for lens modeling \citep[see, e.g.,][]{keeton01b} and commonly used lens parameterizations but is independent of the analysis by \citet{kochanek_dalal04}, suggesting that parity dependence is a generic result of parametric lens modeling.  In addition, the errors in the fluxes (from less than one percent to 10\%) are not large enough to obliterate the observed effect.  

\begin{table}
\caption{Best-Fit Parameters for Observational Sample}
\label{tab:DK02}
\begin{tabular}{llllr}
\hline\hline
Lens & $b ~(\arcsec)$ & $q$  & $\gamma$& $\theta_{\gamma}$ (rad) $^a$\\
\hline
MG0414+0534 & 1.37 & 0.75 & 0.11 & 0.98 ~~~~~~\\
B0712+472 & 1.05 & 0.46 & 0.10 & 1.73  ~~~~~~\\
B1422+231 & 0.94 & 0.69 & 0.16 & 0.07 ~~~~~~\\
B1933+503 & 0.67 & 0.60 & 0.06 & $-$0.22 ~~~~~~\\
B2045+265 & 1.36 & 0.63 & 0.06 & 0.62 ~~~~~~\\
\hline
\end{tabular}
\\
$^a$ Shear orientation is relative to halo main axis.
\end{table}  

The best-fit parameters for $b$, $q$, $\gamma$, and $\theta_{\gamma}$ are shown in Table \ref{tab:DK02}.  We use these parameters, in addition to the halo orientation and the best-fit source and lens positions to create a set of smooth models on which to test substructure models.  

\section{Mock Lenses}
\label{sec:mock}

\subsection{Generating Mock Observations}
\label{sec:mock_obs}

We test substructure models with mock observations of lens systems.  We begin with a set of models for lens halos and lens environments:  the best-fit smooth macromodels of the previous section as summarized in Table \ref{tab:DK02}.  To the lens halos we add clumps as discussed in the following section. 

For a single mock observation of a lens macromodel and substructure, we choose a source position and find the associated image positions and magnifications by solving the lens equation using a Newton-Raphson method on a grid of possible image positions.  Gaussian observational errors are added to image positions and lens galaxy position:  we adopt an observational error of 3 mas.   Systems that result in 4 ``observed" images are modeled using our automated lensing code, and the results are used to create a cumulative magnification perturbation distribution.

When we choose a source position, we have two options.  We can use the best-fit source positions of the observational sample.  This results in a lens configuration (e.g., cusp, fold, or cross) that is similar to that of the observed lenses.  Alternately, we can sample the source plane and create a variety of lens configurations.  When sampling the source plane, we account for magnification bias, assigning source positions by sampling the image plane uniformly as described by \citet{keeton_zabludoff04}.\footnote{\citet{keeton_zabludoff04} show  that for sources with a power law
luminosity function, $dN/dS \propto S^{-\nu}$  with $\nu=2$,
magnification weighting in the source plane is equivalent  to uniform
sampling of the {\it image} plane.}  Results where the source plane is sampled with a uniform weighting in the image plane are labeled `UW.'   Regardless of how the source positions are assigned, we create and model $\sim$1000 realizations for each lens macromodel, for a total of $\sim$5000 realizations.

\subsection{Substructure Models}

In order to approximate the clumpy distributions of matter we expect to find in galaxy halos, we add dark matter substructure to the smooth macromodel.  These clumps are modeled as projected Moore-like profiles (see, for comparison, Moore et al. 1999) with density profile
\begin{equation}
\kappa(x) = \left\{ 
\begin{array}{l l}
  \kappa_{\rm s}(x^{1/2}+x^2)^{-1} &~, R \leq r_{\rm t}  \\
 0&  ~ ,R > r_{\rm t} \\
\end{array} \right. 
\end{equation}
where $x = R/r_{\rm s}$, $R$ is the projected separation from the center of the clump, $r_{\rm s}$ is the scale radius, and $r_{\rm t}$ is the tidal radius of the clump.  We test substructure models based on simulations and ad hoc substructure models as described in the following sections.    

\subsubsection{\citet{mao_etal04} Substructure Model}
\label{sec:mao_descrip}

Lensing is most sensitive to the clumps near the Einstein radius of the lens galaxy.  \citet{mao_etal04} estimate that the Einstein radius is  $\sim$3\% of the virial radius of a lensing halo ($b=0.03R_{\rm vir}$).   If we assume that the halos used in the simulations of \citet{mao_etal04} can be modeled by SIEs, then the mass within $b$ is also $\sim$3\% of the halo mass ($M(<b)=0.03M_{\rm halo}$).  Given a surface mass density in subclumps at the Einstein radius of $\sim$0.5\%, the total mass in subclumps is $\sim10^{-4} M_{\rm halo}$.  The mass resolution of the simulations of \citet{mao_etal04} is also $10^{-4} M_{\rm halo}$, and simulations have shown that smaller clumps are significantly more common than larger clumps.  We can then assume that the result of \citet{mao_etal04} can be approximated by a substructure model where 1 clump of $10^{-4} M_{\rm halo}$ is centered on the Einstein radius.  

\citet{mao_etal04} also find that most clumps have a tidal radius which is $\sim$0.1 of the halo virial radius.  Thus, clumps that are found at separations up to four times the Einstein radius, $4b$, contribute to the measurable substructure mass density.  A better substructure model would include subclumps  within $4b$ of the lens center.  Within that radius, \citet{mao_etal04} find a substructure mass fraction of $\sim$2\%.   Our fiducial substructure model, then, includes 24 subclumps of $10^{-4}M_{\rm halo}$ within $4b$ of the lens center.  Each clump has a tidal radius equal to 0.1 of the halo virial radius.  Fiducially, we set the scale radius to $b/20$.  The clumps are placed to follow the SIE density distribution.  We refer to this model subsequently as M04.  

\subsubsection{The `ALL' Substructure Model}

As the substructure model based on numerical simulations will prove insufficient to reproduce the observed parity dependence, we create a substructure model with greater effect on the magnification perturbation.  This model places one clump within a scale radius of each of the positions where images {\it would be found in the absence of substructure}.  We refer to this substructure model as `ALL' since  all images have a corresponding clump.  Fiducially, we use the same tidal radius, mass, and scale radius as in the M04 model.

\section{Results}
\label{sec:results}

When we model real observations in Section \ref{sec:observations}, we are obliged to compare observed fluxes to modeled fluxes.  In our mock observations, however, we can `observe' magnifications and, thus,  we compare observed and modeled magnifications.  The magnification perturbation is estimated by
\begin{equation}
\delta_i= {\rm log} (\mu_{{\rm obs,}i}/\mu_{{\rm mod,}i}),
\end{equation}
where $\mu_{{\rm obs,} i}$ is the observed magnification of the $i$th image and $\mu_{{\rm mod,} i}$ is the modeled magnification of the $i$th image

We estimate the magnification perturbations for three different scenarios for substructure in the following section.  First, we see if substructure models based upon dark matter simulations are sufficient to induce parity dependence.  Second, we investigate whether satellite galaxies in lens halos can account for the observed parity dependence.  Finally, we design a substructure model that does induce parity dependence and discuss the implications of such a model.

\subsection{Comparison to \citet{mao_etal04} Substructure Model}
\label{sec:m04_results}

We investigate if the substructure constraints found in the simulations of \citet{mao_etal04} can reproduce the observed parity dependence in this section.  We first test the fiducial M04 model described in Section \ref{sec:mao_descrip}.  Then, explore the effects of varying substructure model parameters individually:  $r_{\rm s}$, $r_{\rm t}$, subclump mass, and $f_{\rm sub}$.  

\begin{figure}[h]
\centering
\resizebox{3.in}{!}
	{\includegraphics{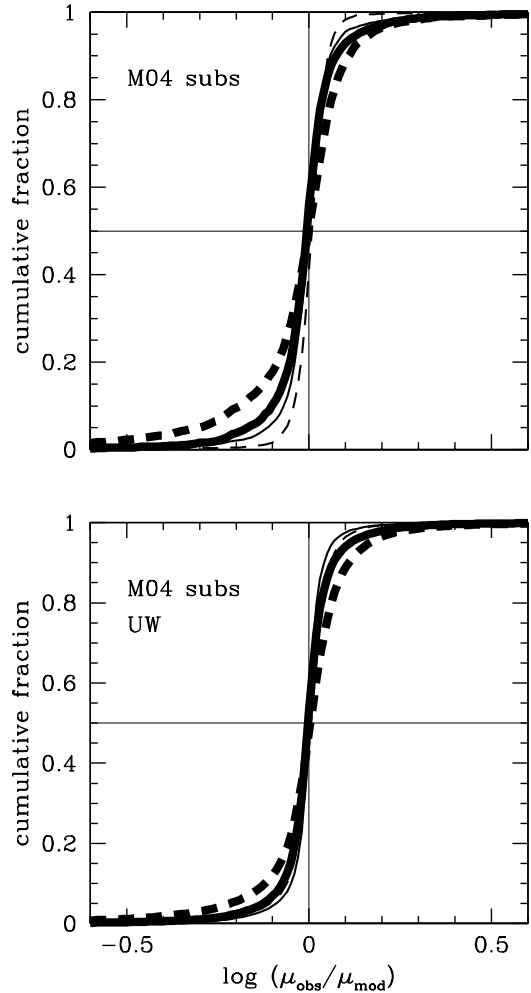}}
\caption{Parity dependence in the cumulative distribution of magnification perturbations using the substructure model following \citet{mao_etal04}.  The top panel shows the observed lens configurations  while the bottom panel varies the source positions used (see text).  Solid lines represent positive parity images, while dashed represent negative parity images.  The brighter of two images with the same parity is denoted by a thick line and the fainter by a thin line. \label{fig:m04}}
\end{figure}

The cumulative distribution of magnification perturbations, $N(<\delta_i)/N_{\rm tot}$, using the fiducial M04 model are plotted in Figure \ref{fig:m04}.  Here, we can see results that appear to be consistent with \citet{rozo_etal06}:  the median perturbation is near zero, although the tails are not negligible.  
  
Given the small size of the observational sample, the exceptionality of the observed parity dependence is best quantified by estimating the probability of finding a sample of 5 lenses with significant parity dependence.  Increasing the number of realizations modeled such that each lens macromodel is used at least 5000 times, we draw 5000 sets of 5 lenses (one mock observation for each of the lens macromodels).  We create test criteria for observing parity dependence in the sets.  One important aspect in developing criteria is specifying how different the magnification perturbations of $\mathcal I_{-\bigstar}$ and $\mathcal I_{+\bigstar}$ are in order to exclude sets where the difference is negligible and no parity dependence would be observed.  $\delta_{f,+\bigstar} - ~\delta_{f,-\bigstar}  = 0.06$ is the smallest difference seen in the observational sample, and we adopt this in the criteria for $\delta_{+\bigstar} - ~\delta_{-\bigstar}$.  The three sets of criteria we test are: 
\begin{enumerate}
\item $\delta_{-\bigstar} < \delta_{+\bigstar}$ everywhere and at least 60\% of the distribution shows  $\delta_{+\bigstar} - ~\delta_{-\bigstar} \ge 0.06$\\
\item $\delta_{-\bigstar} < \delta_{\rm all ~others}$ everywhere and at least 60\% of the distribution shows  $\delta_{+\bigstar} - ~\delta_{-\bigstar} \ge 0.06$\\
\item $\delta_{-\bigstar} < \delta_{\rm all ~others}$ and $\delta_{+\bigstar} - ~\delta_{-\bigstar} \ge 0.06$ everywhere (``{\it parity dependence probability}")
\end{enumerate}
Of the 5000 sets, 170 (3\%) satisfy the first set of criteria, 58 sets (1\%) satisfy the second, and 14 sets (0.3\%) satisfy the third.  So, given the M04 substructure model, the probability of finding a parity dependence like the one observed is small but maybe not insignificant.  We designate the third set of criteria  as {\it the parity dependence probability}.  This measure is more useful than a standard Kolmogorov-Smirnov test, as we are most interested in the relation between magnification perturbations of different images, not comparing the absolute distribution of observed flux perturbations to magnification perturbations of mock observations.  

In the bottom panel of Figure \ref{fig:m04}, we vary the source positions as described in Section \ref{sec:mock_obs}.  Here, the parity dependence probability is 5 out of 5000 sets (0.1\%).  The particular lens configurations in the observed sample, then, seem to be mildly more effective in creating parity dependence than a sampling of source positions.  Henceforth, we use a smaller number of sets to calculate the probability, a total of 500.  

\begin{figure}[h]
\centering
\resizebox{3.3in}{!}
	{\includegraphics{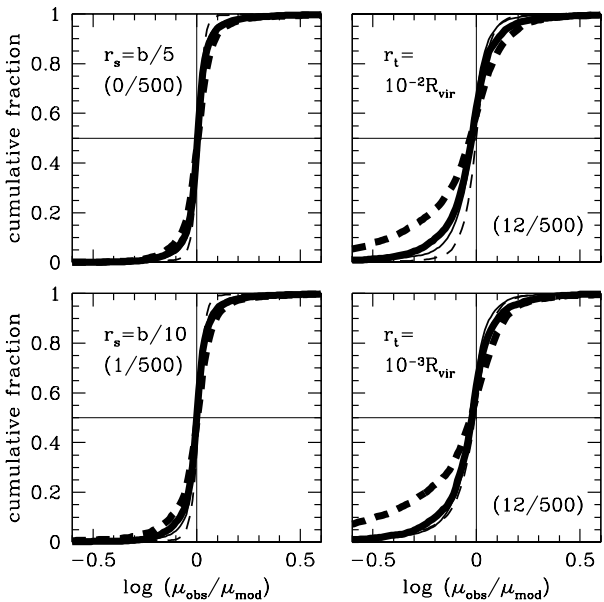}}
\caption{Varying the scale radius and tidal radius.  The left-hand panels test larger values of the scale radius while the right-hand panels test smaller values of the tidal radius.  The parity dependence probability for each test is in parentheses.  Solid lines represent positive parity images, while dashed represent negative parity images.  The brighter of two images with the same parity is denoted by a thick line and the fainter by a thin line.  \label{fig:rtrs}}
\end{figure}

\begin{table}
\caption{Results Using Substructure Models Based on \citet{mao_etal04}}
\label{tab:m04}
\begin{tabular}{lllllrl}
\hline\hline
$f_{\rm sub} (< 4b)$ & $N$ & $M/M_{\rm halo}$  & $r_{s}/b$& $r_{t}/R_{\rm vir}$ & probability \\
\hline
0.02 & 24 & $10^{-4}$ & 0.05 & 0.1 & 1/500&$^a$ \\
0.02 & 24 & $10^{-4}$ & 0.2 & 0.1 &  0/500 \\
0.02 & 24 & $10^{-4}$ & 0.05 & 0.01 & 12/500 \\
0.25 & 300 & $10^{-4}$ & 0.05 & 0.1 &  10/500\\
0.25 & 30 & $10^{-3}$ & 0.05 & 0.1 & 16/500 \\
0.25 & 3000 & $10^{-5}$ & 0.05 & 0.1 &  0/500\\
0.02 & --- &---  &--- & ---& 1/500&$^b$ \\
\hline
\end{tabular}
\\
$^a$ fiducial M04\\
$^b$ $n \propto m^{-1.8}$ mass function
\end{table}  

We have chosen to parameterize the substructure model of \citet{mao_etal04} in a reasonable, but not unique, way.  Here we test how the parity dependence depends on the parameters chosen, and results are summarized in Table \ref{tab:m04}.  We vary the tidal radius $r_t$ or the scale radius $r_{s}$ individually, while keeping the other subclump parameters fixed at fiducial values in Figure \ref{fig:rtrs}.  Increasing the scale radius reduces the parity dependence induced by substructure, as the mass within the scale radius is spread out over a larger area.  Decreasing the tidal radius puts more mass into a smaller area and increases the parity dependence induced by substructure.  The parity dependence probability increases by an order of magnitude to $\sim$ 2\%,  but remains small.  Varying the scale radius or the tidal radius does not change the parity dependence probability in a linear way.  For example, the smallest tidal radius, $10^{-3}R_{\rm vir}$, is smaller than the fiducial scale radius and has the same parity dependence probability as a substructure model with tidal radius of $10^{-2}R_{\rm vir}$.  

We do an additional test, modeling the subclumps as SISs which has steeper inner profiles than our Moore-like profiles and which might better approximate clumps with satellite galaxies in their centers.  At fixed mass, however, the perturbers are no better than the M04 in inducing parity dependence.  

\begin{figure}[h]
\centering
\resizebox{3.in}{!}
	{\includegraphics{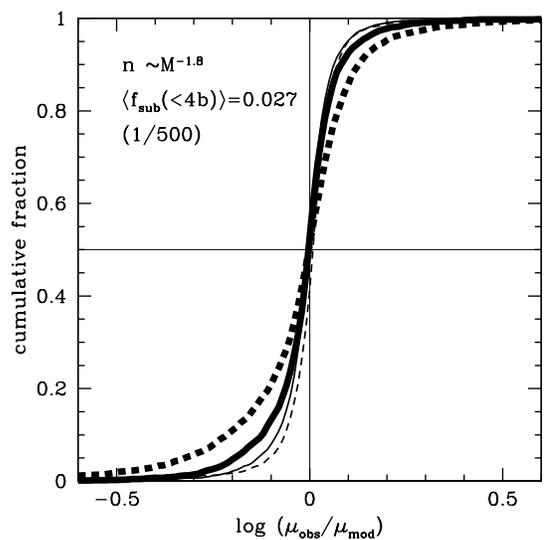}}
\caption{Test using a realistic mass function.    The parity dependence probability is in parentheses.  Solid lines represent positive parity images, while dashed represent negative parity images.  The brighter of two images with the same parity is denoted by a thick line and the fainter by a thin line.  \label{fig:mf}}
\end{figure}  

We try a slightly more realistic version of the substructure model, including a mass function for subclumps.  Dark matter simulations suggest that  $n(m) \sim m^{-1.8}$ \citep{ghigna_etal00}.  The scale radius is scaled by the square root of the mass:  at $M=10^{-4}M_{\rm halo}$, $r_{\rm s}$ has the value used in the fiducial model, and it is larger for larger masses and smaller for smaller masses.  The results are shown in Figure \ref{fig:mf}, where it is clear that this more realistic model is no more effective in inducing parity dependence than the fiducial model.  

\begin{figure}[h]
\centering
\resizebox{2.5in}{!}
	{\includegraphics{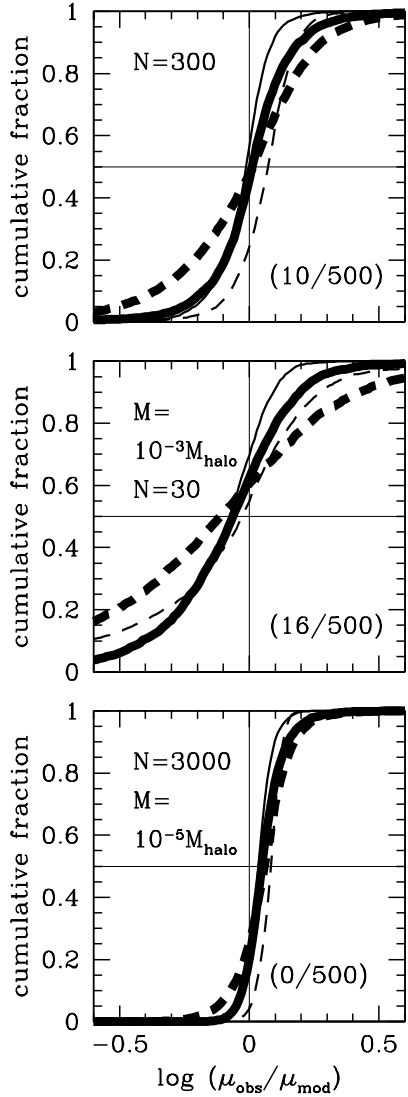}}
\caption{Testing a larger substructure fraction ($f_{\rm sub} = 25\%$ within $4b$) and different substructure masses:  fiducial mass clumps ({\it top}), $10^{-3} M_{\rm halo}$ mass clumps ({\it center}), and $10^{-5} M_{\rm halo}$ mass clumps ({\it bottom}).  The parity dependence probability for each test is in parentheses.  Solid lines represent positive parity images, while dashed represent negative parity images.  The brighter of two images with the same parity is denoted by a thick line and the fainter by a thin line.  \label{fig:fsub}}
\end{figure}

The amount of substructure in the M04 models seem insufficient to match the observational results, unless our observational results represent an outlier in the distribution of magnification perturbations.  The M04 model, however, represents the substructure distribution at $z=0$.  Since a significant number of strong lenses are found nearer to $z=1$, a significantly larger amount of substructure is expected.  If we increase the substructure mass fraction, the size of the magnification perturbations correspondingly increases, as does the probability of observed parity dependence.  This can be seen in Figure \ref{fig:fsub}, where 300 clumps are used (which are 12.5 times more than in the fiducial case and correspond to a substructure mass fraction within $4b$ of 25\%).  Using a larger substructure mass fraction, we also test the effect of substructure mass.  Keeping both the tidal radius and scale radius fixed, fewer, larger mass clumps are more effective than a greater number of smaller mass clumps, but the parity dependence probability remains no greater than a few percent.  

Substructure models based on dark matter simulations, then, seem insufficient to account for the observed parity dependence unless the observed parity dependence is an outlier in the distribution.  If, however, the substructure mass fraction is an order of magnitude larger, the probability of the observed parity dependence may be a few percent.  In addition, clumps that are well- concentrated have a greater effect than those that are less concentrated.  Well-concentrated clump might be found in higher resolution dark matter simulations:  clumps that are below the mass resolution and/or are tidally destroyed faster in lower resolution simulations.  In addition, in simulations that include the effects of baryons, we may see longer surviving subclumps that have their central densities boosted by cold baryonic components.   

\subsection{The Effect of Luminous Satellites}

\begin{figure}[h]
\centering
\resizebox{3.in}{!}
	{\includegraphics{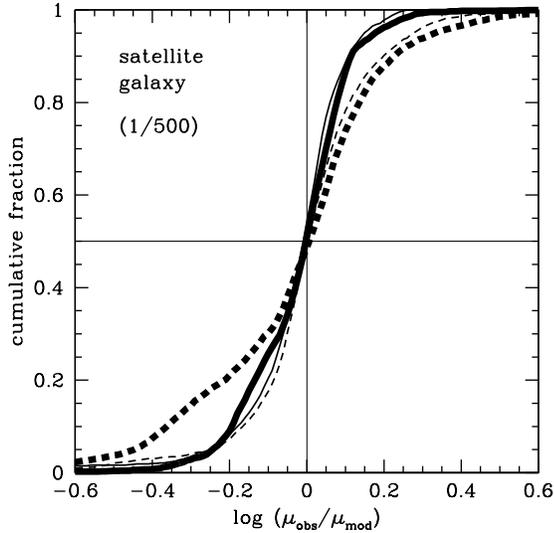}}
\caption{Testing the effect of a single luminous satellite.   The parity dependence probability is in parentheses.  Solid lines represent positive parity images, while dashed represent negative parity images.  The brighter of two images with the same parity is denoted by a thick line and the fainter by a thin line.  \label{fig:lumsat}}
\end{figure}

We find it difficult to reproduce the observed parity dependence using realistic models of dark matter substructure.  However, of the 5 lenses in our sample, 2 of them (MG0414+0534, B2045+503) have luminous companions that lie in and around the Einstein radius of the lens halo.  These companions may be satellite galaxies in the lens plane.\footnote{The prevalence of luminous satellites in anomalous flux ratio systems has been noted previously:  5 of the 22 gravitational lenses found in the Cosmic Lens All-Sky Survey \citep[CLASS;][]{browne_etal03,myers_etal03} have luminous companions (B1608+656, B2045+503, MG0414+0534, B1127+385, B1359+154).}  We test the effect of placing a single luminous satellite within two Einstein radii of the lens center in Figure \ref{fig:lumsat}.  The satellite is modeled as a SIS with truncation radius set to the Einstein radius of the lens halo $(r_{\rm t, sat}=b)$.  The Einstein radius of the SIS is one-tenth the Einstein radius of the lens $(b_{\rm sat}=b/10$) and it has a mass of 0.01$M_{\rm halo}$.  This is similar to the best-fit lens models for the companion objects in MG0414+0534 and B2045+503 \citep{ros_etal00,mckean_etal07} which parameterize the companions as untruncated SISs and have best-fit values of $b_{\rm sat} = 0.17$ and 0.08, respectively.  In our case, while the luminous satellite seems to increase the absolute size of the magnification perturbations, the probability of observing parity dependence is estimated to be as low as in the M04 substructure model (1 out of 500 sets of realizations).  

If the additional mass were much larger, it could 1.) steepen the inner profile of the lens from a SIE and 2.) significantly increase the Einstein radius of the lens.  This would induce a parity dependence such that all $\delta_i < 0$ and $\delta_{-\bigstar} < ~\delta_{+\bigstar}$.  However, faint negative parity images would be just as or more demagnified compared to bright negative parity images.

In a statistical sense, luminous satellites do not appear to resolve the observed parity dependence.  Regardless, we additionally check to see if they could eliminate the observed parity dependence in the specific cases of MG0414+0534 and B2045+503.  

\begin{table}[h]
\caption{Flux Perturbations in MG0414+0534 and B2045+503}
\label{tab:lumsat}
\begin{tabular}{lllll}
\hline\hline 
Lens and Model & $\delta_{f,+\bigstar}$ & $\delta_{f,+\ast}$ & $\delta_{f,-\bigstar}$ & $\delta_{f,-\ast}$ \\
\hline
\noalign{\smallskip}
MG0414+0534 & -0.09 & 0.11 & -0.16 & 0.07\\
~~~ include sat & 0.013 & 0.12 & -0.052 & -0.013\\
B2045+503  &-0.18 & -0.12 & -0.53 & 0.57 \\
~~~ include sat & -0.27 & 0.16 & -0.73 & 0.70 \\
\noalign{\smallskip}
\hline
\end{tabular}
\end{table}

MG0414+0534 exhibits a fold configuration -- two of the four images are found close to each other.  The source and lens redshifts are found at $z_{\rm s}=2.6$ and $z_{\rm l}=0.96$, respectively \citep{lawrence_etal95,tonry_kochanek99}.  The companion object lies near the Einstein radius and far from the fold images.  It does not have a measured redshift, but we assume that it lies in the lens plane.  We find an initial smooth lens model, using 10 constraints and 9 parameters.  We add the companion object to the lens model as a SIS, fixing its position to match observations and its Einstein radius to 0.12 times the Einstein radius of the lens.\footnote{\citet{ros_etal00} find a larger value of the Einstein radius of the satellite than we use.  We find that larger values for the Einstein radius induce extra images in the lens.  Note also that we use the fluxes from \citet{katz_etal97} and not those in the \citep{ros_etal00}, although results are consistent.}  We add no new parameters nor constraints and refit, finding a new, smaller $\chi^2$.  The magnification perturbations for each image are shown in Table \ref{tab:lumsat}.  Here the inclusion of the satellite object clearly reduces the size of the magnification perturbations.  While in the original fit, all magnification perturbations were $\sim$25\%, now two of the four are less than 5\%.  However, the size of $\delta_{f,+\bigstar} - ~\delta_{f,-\ast}$ remains consistent.  In addition, $\delta_{f, -\bigstar} < \delta_{f, \rm all~others}$.  This parity dependence appears robust:  all observed fluxes show that $\mathcal I_{+\bigstar}$ is brighter than $\mathcal I_{-\ast}$, while all best-fit modeled magnifications find that $\mu_{-\bigstar} > ~\mu_{+\bigstar}$.  The brightest images in the system are the fold images, and, given the distance of the satellite object, it is not surprising that it does not seem to effect the ratio of those fluxes.  
 
B2045+503 exhibits a cusp configuration with three images positioned near each other.  Its source and lens redshifts have been measured at $z_{\rm s}=1.28$ and $z_{\rm l}=0.867$, respectively \citep{fassnacht_etal99}.  The companion galaxy (G2)  lies between the lensing galaxy (G1) and the cusp images and has a photometric redshift consistent with the lensing galaxy \citep{mckean_etal07}.  The best-fit smooth model by \citet{mckean_etal07} for G1 and G2 finds that the ellipticity of G1 is oriented 90 degrees away from the cusp images while a large external shear is pointed along the line from G1 to the cusp images.  In addition, their best-fit model for G2 is very elliptical ($q=0.133$) and pointed along the same direction as the shear.  Including the previously found best-fit parameters for G2 --  Einstein radius, axis ratio, orientation, and position -- we can significantly reduce the $\chi^2$ of our original best-fit, but the magnification perturbations are not reduced in this case:  $\delta_{f,-\bigstar} < 0$ and $\delta_{f,-\bigstar} < ~\delta_{f,\rm ~all ~others}$.  As in the case of MG0414+034, the observed fluxes show that $\mathcal I_{+\bigstar}$ is brighter than $\mathcal I_{-\ast}$, while the best-fit modeled magnifications find that $\mu_{-\bigstar} > ~\mu_{+\bigstar}$.  We use the observed fluxes from \citet{mckean_etal07}.  \citet{koopmans_etal03} find that B2045+503 is affected by variability in the radio fluxes, however, our results are relatively unaffected. 

Luminous satellites, then, do not appear sufficient to explain the observed parity dependence either in the statistical sense or in the particular cases in our observational sample.  

\subsection{Creating an Effective Substructure Model ('ALL')}
\label{sec:all}

Given the difficulty in reproducing the observed parity dependence with well-motivated models of dark and luminous substructure, a pertinent question to ask is what substructure model will induce parity dependence.  There exists a subsample of M04 substructure realizations which do show appreciable parity dependence -- the fraction of those realizations with at least one clump located within a scale radius of an observed image.  This sample has a parity dependence probability of $\sim$14\%;  however, the fraction of realizations in which the M04 substructure model results in a close alignment of clump and image is small, $\sim4\%$. 

Based on that result, we create a substructure model with greater effect on the magnification perturbation.  This model places one clump within a scale radius of each of the image positions that {\it would be found in the absence of substructure}.  As previously noted, we refer to this substructure model as `ALL'  because all images have a corresponding clump.

\begin{figure}[h]
\centering
\resizebox{3.in}{!}
	{\includegraphics{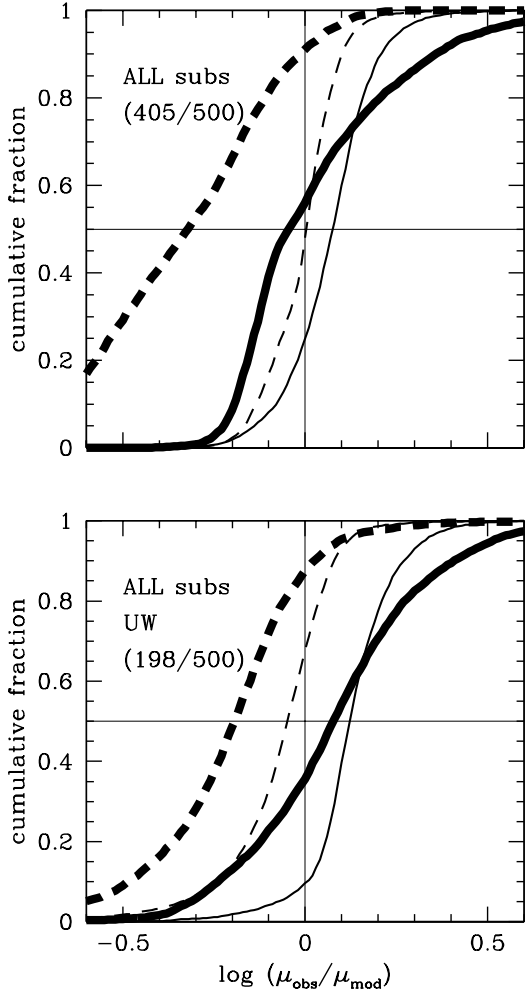}}
\caption{Parity dependence when each image in a lens is paired with a nearby subclump.  The top panel shows results using the best-fit source positions from the original data, while the bottom panel samples the source plane.   The parity dependence probability is in parentheses.  Solid lines represent positive parity images, while dashed represent negative parity images.  The brighter of two images with the same parity is denoted by a thick line and the fainter by a thin line.  \label{fig:all}}
\end{figure}

The results featuring a clear parity dependence are shown in Figure \ref{fig:all}.  Qualitatively, the distributions are similar to the results of \citet{kochanek_dalal04}:  $\delta_{-\bigstar} < ~\delta_{+\bigstar}$ while the distribution of $\delta_{+\bigstar}$ and $\delta_{\rm faint~images}$ are very similar.  The UW test has similar results although the faint images do not correspond to $\delta_{+\bigstar}$ as in the previous case.   The probability of observing parity dependence has increased significantly to 40\% and 80\% when sampling the source plane and when using specific source positions, respectively.  A similar model -- placing small, truncated SISs near images -- has been tested by \citet{kochanek_dalal04} with similar qualitative results.  

The efficacy of this model suggest that the requirements for parity dependence may be significant numbers of well-concentrated {\it and} well-aligned perturbers.  It also supports the possibility that more substructures than the amount found by \citet{mao_etal04} are required to achieve the observed parity dependence.  While the ALL substructure model uses fewer subclumps and, therefore, has a smaller measurable substructure mass fraction (1/6th of the \citet{mao_etal04} result), the placement of clumps is perverse.  We expect that, in a realistic substructure model, the placement of clumps in the projected plane is stochastic, and the probability of a chance alignment between image and clump is small.  

\begin{figure}[h]
\centering
\resizebox{3.3in}{!}
	{\includegraphics{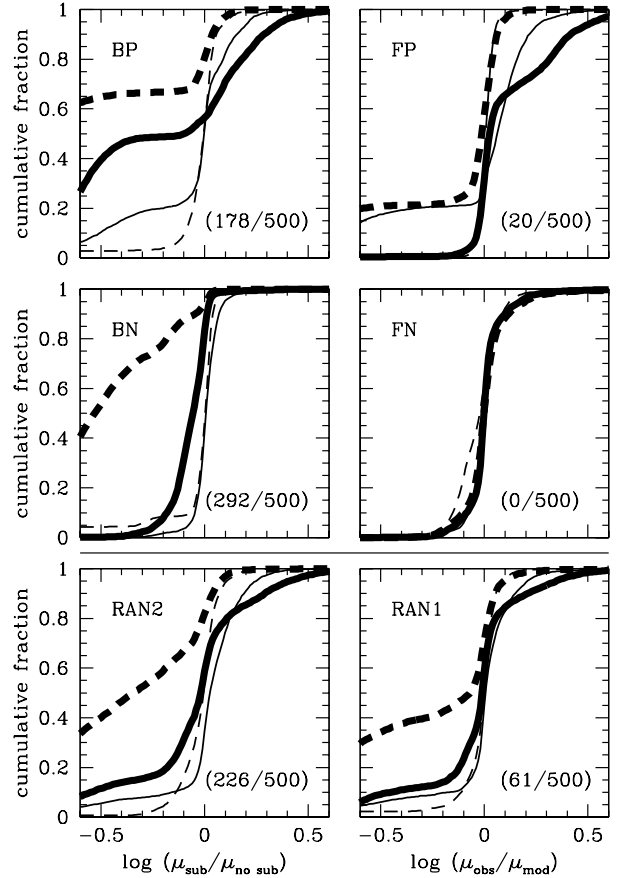}}
\caption{Testing the effect of individual clumps on the parity dependence.  For the top 4 panels, clockwise from the top-left, each panel shows the results for placing one clump only near the bright positive parity image (BP), faint positive parity image (FP), faint negative parity images (FN) and bright negative parity image (BN).  The bottom-right panel shows the effect of placing one clump near a random image position and the bottom-left panel shows the effect of placing two clumps near two random image positions.    The parity dependence probability is in parentheses.  Solid lines represent positive parity images, while dashed represent negative parity images.  The brighter of two images with the same parity is denoted by a thick line and the fainter by a thin line.  \label{fig:testim}}
\end{figure}

Now that we have a substructure model that induces the observed parity dependence, we test which of the clumps are the most effective in Figure \ref{fig:testim}.  Clumps near faint images have relatively mild effects;  clumps near bright images show larger effects on the magnification perturbation distribution.  In particular, clumps near $\mathcal I_{-\bigstar}$ cause a significant demagnification in that image, while clumps near $\mathcal I_{+\bigstar}$ cause large negative magnification perturbations in that image, as well as a pronounced large demagnification tail in $\mathcal I_{-\bigstar}$.  Intriguingly, although at least some of the lens configurations have $\mathcal I_{+\bigstar}$ and $\mathcal I_{-\bigstar}$  as a close pair, clumps near $\mathcal I_{-\bigstar}$ have little effect on $\delta_{+\bigstar}$, while the converse is not true for clumps near $\mathcal I_{+\bigstar}$.  In the bottom panels of Fig. \ref{fig:testim}, one chance alignment between image and clump is sufficient to boost the probability of observing parity dependence, and two chance alignments puts the probability at nearly 50\%.  

The results of this model can be described simply by inspecting the eigenvalues of the magnification matrix.  The lens equation describes the relation between the source and image positions ${\mathbf u} = {\mathbf x} - \nabla \psi ({\mathbf x})$, where $\mathbf u$ is the source position, $\mathbf x $ is the image position, and $\psi$ is the lensing potential.  The magnification matrix is given by $\partial {\mathbf u}/\partial {\mathbf x}$.  The eigenvalues of the magnification matrix are given by 
\begin{equation}
\lambda_1 = 1- \kappa - \gamma
\end{equation}
and
\begin{equation}
\lambda_2 = 1- \kappa + \gamma,
\end{equation}
and the magnification is $\mu = 1/(\lambda_1 \lambda_2)$.  In the case of double-positive parity images, both $\lambda_1 > 0$ and $\lambda_2 > 0$, while in the case of negative parity images $\lambda_1 < 0$ and $\lambda_2 > 0$.  In the case of the brightest images -- for example, a fold -- we would expect that $\lambda_1 \sim 0$ for both images.  By placing a clump near the image, we add a little to the surface mass density $\kappa$.  For the positive parity image, in general, this makes $\lambda_1$ smaller and closer to zero and increases the magnification.  For the negative parity image, in general, this makes $\lambda_1$ more negative and further from zero, decreasing the magnification.  Thus, generically, adding mass near images should make 1.) the brightest negative parity image dimmer than a smooth model would suggest ($\delta_{-\bigstar} < 0$) and 2.) make the brightest positive parity image brighter than a smooth model would suggest ($\delta_{+\bigstar} > 0$).  

\begin{figure}[h]
\centering
\resizebox{3.in}{!}
	{\includegraphics{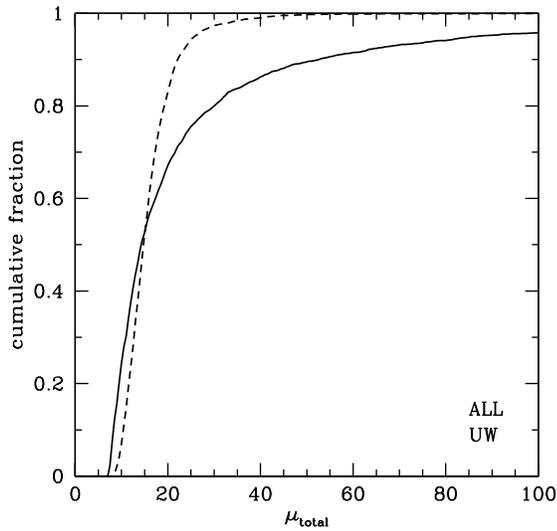}}
\caption{The cumulative distribution of the total magnification for realizations using the ALL substructure model and sampling the source positions (dashed line).  The solid line shows the distribution of total magnification using no substructure.  \label{fig:magtot}}
\end{figure}

As previously noted, a substructure model such as ALL is not plausible.  But it might be realistic if it induced a large increase in the lensing probability.  Two possibilities are that clumps have increased the size of the tangential caustic.  Given that the mass fraction in substructures added by the ALL model is very small, this possibility is unlikely.  The other possibility is that the total magnification of the lensed systems is increased making the lens system more likely to be observed.  This seems highly unlikely given that the main effect is to demagnify one image greatly.  Nonetheless we test this possibility using the ALL substructure model and sampling the source positions in Figure \ref{fig:magtot}, where we plot the cumulative distribution of total magnification.  At the smallest magnifications, the ALL substructure model has larger total magnification than in the no substructure case.  This is probably the result of the lenses having more mass in the ALL case than in the no substructure case and does not reflect any real magnification bias.  The cumulative distribution in the ALL case rises more steeply and, overall, the result is that substructure actually decreases the total magnification.

\section{Discussion}
\label{sec:discussion}

We discuss here the different contributions to the observed parity dependence from 1.) lens modeling and 2.) astrometric perturbations.

\begin{figure}[h]
\centering
\resizebox{3.3in}{!}
	{\includegraphics{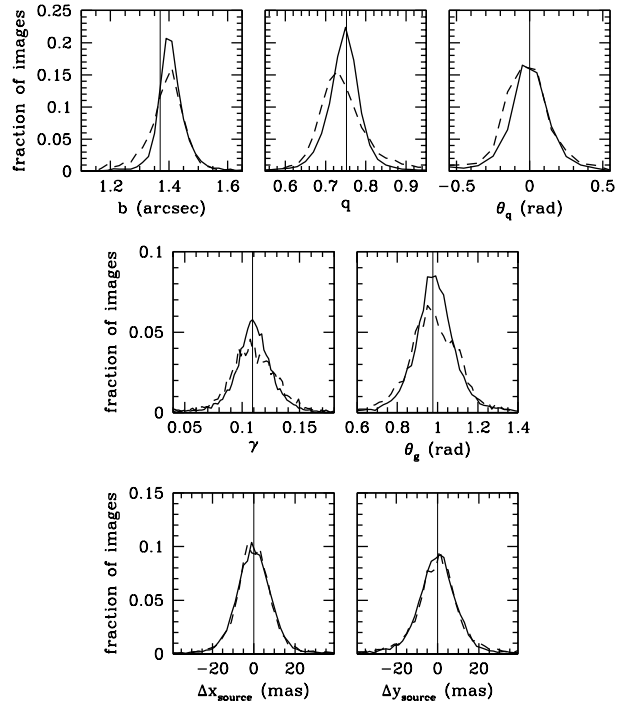}}
\caption{The best-fit smooth model parameters using M04 (solid lines) and ALL substructure models (dashed lines) in conjunction with the best-fit macromodel for MG0414+0534 (see Table \ref{tab:DK02}).  Source points are sampled.  The input macromodel values are shown as vertical lines.  \label{fig:param}}
\end{figure}

In general, we find lens modeling finds appropriate best-fit parameters (see Figure \ref{fig:param}).  The distribution of best-fit parameters peaks around the no-substructure input parameter.\footnote{The distributions for the Einstein radius $b$ peak slightly larger than the input value, which is not unexpected given that we added mass to the lenses in the form of clumps.}  In addition, the width of the best-fit parameter values from the M04 and from the ALL substructure models are similar.  The widths of the distribution of best-fits, however, can be fairly large (10-20\%).  A bias in the lens modeling code does not appear to be a contributing factor to the parity dependence.  

\begin{figure}[h]
\centering
\resizebox{3.3in}{!}
	{\includegraphics{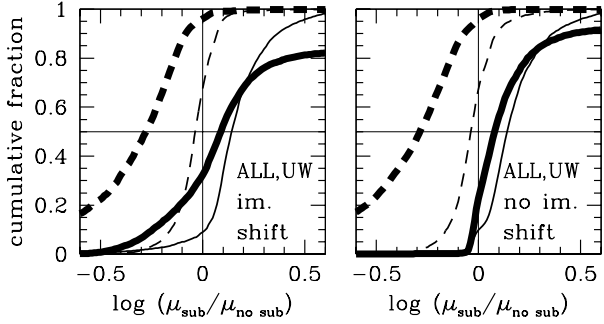}}
\caption{Comparing the magnification of images with and without substructure (no lens modeling).  In the substructure case, each image in a lens is paired with a nearby subclump.   {\it Left}:  The magnification at the image position in a clumpy lens compared to the magnification at the image position of a smooth lens.  {\it Right}: The magnification at the image position in a clumpy lens compared to the magnification at the same image position but using a smooth lens model.  Solid lines represent positive parity images, while dashed represent negative parity images.  The brighter of two images with the same parity is denoted by a thick line and the fainter by a thin line. \label{fig:nofit}}
\end{figure}

In fact, lens modeling as a whole has only a small effect on the parity dependence.  In Figure \ref{fig:nofit}, we perform two tests on the magnifications using no lens modeling and sampling the source positions.  In the left-hand panel, we show the distribution of magnification perturbations using the ALL substructure model with no lens modeling at all.  Instead of comparing the observed and modeled image magnifications, we compare the magnification at the image position of a mock lens with substructure with the magnification at the image position of a similar mock lens but without substructure.  This result includes the effect of shifts in the image positions due to substructure:
\begin{equation}
\delta_{\rm im. ~shift} = {\rm log} \left(\frac{\mu_{\rm obs,sub}(\vec{x_{\rm obs,sub}})}{\mu_{\rm obs,no~sub}(\vec{x_{\rm obs,no~sub}})}\right)
\end{equation}  
In the right-hand panel, we  compare the magnification at the image position of a mock lens with substructure with the calculated magnification at the same image position (the image position given substructure) but putting no substructure into the mock lens.  This result, then, has no shift in image position.  This test is similar to simple explanation for parity dependence that we described in Section \ref{sec:all}:
\begin{equation}
\delta_{\rm no ~im.~ shift} = {\rm log} \left(\frac{\mu_{\rm obs,sub}(\vec{x_{\rm obs,sub}})}{\mu_{\rm obs,no~sub}(\vec{x_{\rm obs,sub}})} \right)
\end{equation}   In both cases, strong parity dependence of the kind seen in Fig. \ref{fig:all} is evident even though we have done no lensing modeling. 

\begin{figure}[h]
\centering
\resizebox{3.3in}{!}
	{\includegraphics{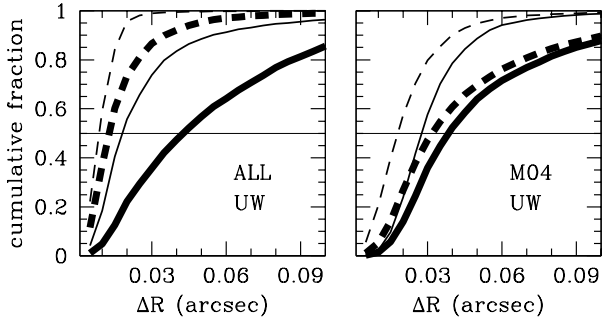}}
\caption{Measuring the astrometric perturbations with no lens modeling..  {\it Left}:  The results for placing one clump near each image (ALL).  {\it Right}:  The results using a substructure model based on \citet{mao_etal04} (M04).   Solid lines represent positive parity images, while dashed represent negative parity images.  The brighter of two images with the same parity is denoted by a thick line and the fainter by a thin line. \label{fig:pos}}
\end{figure}

We examine the size of the shifts in image position due to substructure, and without any lens modeling, in Figure \ref{fig:pos}.  The differences between using the M04 substructure model and the ALL substructure model seem modest.  In both cases, the median change in image position between a lens system with no substructure and one with substructure is $\sim$30 mas.  However, when examining the images separately, we see a significant difference between substructure models.  For the ALL substructure model, the astrometric perturbations for $\mathcal I_{-\bigstar}$ are significantly smaller than that of $\mathcal I_{+\bigstar}$.  In the M04 case, these two cases are similarly large. 

As mentioned previously, we add clumps to the positions where the input macromodel would find images in the absence of substructure.  From Figure \ref{fig:pos}, we see that whether the observed image position is near a clump depends on the image.  For example, in the ALL model, it seems that a significant number of faint images and of $\mathcal I_{-\bigstar}$ may remain well aligned with their perturbers, while the observed $\mathcal I_{+\bigstar}$ may have moved a significant distance relative to the scale radius of the clump.

\section{Conclusions}
\label{sec:conclusions}

Radio observations of quasar sources strongly lensed by galaxy lenses provide information on the mass distribution along the line-of-sight and, as such, have been suggested as probes of galaxy substructure and a small-scale test of CDM.  One simple measure of the subclumps is the substructure mass fraction  at the scale of the Einstein radius in a galaxy lens.  Simulations have predicted that this fraction is $f_{\rm sub}\sim0.5\%$ at $z=0$.  

Previous comparisons of observations to simulations have suggested that there may be more substructure than predicted in simulations with varying levels of discrepancy \citep{dalal_kochanek02,maccio_etal06,amara_etal06}.  \citet{kochanek_dalal04} pointed out that such magnification perturbations did not behave identically in each image and the behavior depended upon the parity (orientation) of the image.  The observed parity dependence they found is such that the brightest negative parity image is demagnified with respect to the magnification predicted by smooth models and has a magnification perturbation smaller than that of the other images ($\delta_{-\bigstar} < 0$ and $\delta_{-\bigstar} < \delta_{\rm all ~others}$).  Here, we use the observed parity dependence in order to extract additional information about the substructure distribution.  

We create mock observations of strong lens systems with clumpy lens halos and compare the distribution of magnification perturbations between the mock and observed samples using an automated lens modeling code.  Overall, our results are consistent with previous results which suggest that there is more substructure in the observations than can be accounted for in dark matter simulations.  Our conclusions are summarized as follows:
\begin{enumerate}
\item Parity dependence in a sample of four-image lenses is found such that the brightest negative parity has a magnification perturbation smaller (less magnified or more demagnified) than that of the other images ( $\delta_{-\bigstar} < \delta_{\rm all ~others}$).\\

\item This effect is difficult to reproduce using substructure models based on dark matter simulations;  the probability of generating a sample with properties like the observed sample is less than 1\%. \\ 

\item While a significant fraction of lens galaxies have been found to have luminous companions, satellite galaxies do not appear sufficient to induce parity dependence.  \\

\item Parity dependence requires that images and clumps are well-aligned and clumps are sufficiently concentrated.  \\

\item More substructure than predicted in simulations may be required, either by resolving smaller but well-concentrated clumps or by increasing the overall substructure mass fraction.
\end{enumerate}

Given the small sample of observed radio lenses, it is difficult to put robust constraints on the size or significance of the observed parity dependence.  In addition, while we approximate the observed parity dependence using simple substructure models, we cannot exclude more complicated models for the lens halos.  We also consider only substructure within the lens halo itself and do not consider the possible parity dependence effects of the significant numbers of clumps that may be found along the line-of-sight.  Previous studies of the importance of line-of-sight clumps have found conflicting results \citep{chen_etal03,metcalf05a,metcalf05b}.  The large amount of line-of-sight structure suggests that this may be an effective way of resolving the parity dependence problem.  However, line-of-sight clumps are less effective than substructure in inducing magnification perturbations and the overall effect of line-of-sight clumps is unclear.

Parity dependence as observed in four-image lenses with radio observations is a strong measure of anomalous fluxes and a constraint on substructure in galaxies.  In this paper, we describe the contradictions between the observed parity dependence and previous theoretical estimates and resolve the discrepancies using modeling of mock observations of lenses.  We attempt to constrain the kinds of substructure models that will reproduce the observed parity dependence.  The substructure models necessary to approximate the observed results is consistent with suggestions that there is more substructure in galaxies than in estimates from simulations.  In addition, we show that it is unlikely that anomalous flux ratios are due solely to observed luminous satellite galaxies.  Future studies of CDM substructure using strong gravitational lenses may require larger samples of observed lenses as well as more physically realistic simulations of dark matter and baryons.

\begin{acknowledgements}
We would like to acknowledge Eduardo Rozo for his indispensable efforts in developing the lens modeling code.  We would like to thank Aleksi Halkola and John McKean for their help and advice on modeling observations of B2045+503;  Peter Schneider, Sherry Suyu, and Jan Hartlap for helpful suggestions;  and Andrey Kravtsov for his support.   
\end{acknowledgements}

\appendix

\section{Lens Modeling Algorithm}
\label{sec:appendix}

Our lens modeling algorithm consists of 7 steps (some of which are repeated):

\begin{enumerate}
\item Fix the value for the Einstein radius to half the maximum separation of images. Set $\gamma$, $\theta_{\gamma}$ to zero.  Test values of $q$ and $\theta_{q}$ on a grid in the source plane.\\

\item Using the best fit value of $q$ and $\theta_{q}$, fit for $b$, $q$, and $\theta_{q}$ in the source plane.\\

\item Using the best fit values of $b$, $q$ and $\theta_{q}$, test values of  $\gamma$ and $\theta_{\gamma}$ on a grid in the source plane. \\

\item Fit for $b$, $q$, $\theta_{q}$, $\gamma$, and $\theta_{\gamma}$ in the source plane.\\

\item Do three iterations of fitting $b$, $q$, $\theta_{q}$, $\gamma$, $\theta_{\gamma}$, $x_{\rm center}$, and $y_{\rm center}$ in the source plane.  Each trial perturbs the image positions slightly.  The set of parameters with the smallest $\chi^2$ are accepted as the initial guess parameters for the next iteration.  If the $\chi^2$ increases, those best-fit parameters are rejected.   \\

\item Do up to 10 iterations of fitting in the image plane.  In each iteration, the image positions are perturbed slightly.  The set of parameters with the smallest $\chi^2$ are accepted as the initial guess parameters for the next iteration.  If the $\chi^2$ increases, those best-fit parameters are rejected.  Iterations are stopped if $\chi^2/N$, the chi-squared per number of constraints, drops below 1.  \\

\item  Check the $\chi^2$ for the best-fit parameters using original, observed image positions.  Recheck modeled image positions for best-fit parameters.  If the wrong number of images or parity violation is found, start again (up to 6 times) and set different values for $b$ and $q$ in Step 3.  

\end{enumerate}

The source plane $\chi^2$ is calculated as in \citet{keeton01b} and reproduced here for convenience:
\begin{equation}
\chi^2_{\rm src} = \sum \delta \vec{u}_i^{T} \cdot \mu_{i}^{T} \cdot S_{i}^{-1} \cdot \mu_{i} \cdot \delta \vec{u}_{i},
\end{equation}
where 
\begin{equation}
S_{i} = R_{i}^{T} \left[   
\begin{array}{cc}
      \sigma_{1,i}^2 & 0 \\
      0 & \sigma_{2,i}^2 \\
    \end{array}
\right] R_{i},
\end{equation}
\begin{equation}
R_i = \left[
\begin{array}{rr}
      -\sin\theta_{\sigma,i} &  \cos\theta_{\sigma,i} \\
      -\cos\theta_{\sigma,i} & -\sin\theta_{\sigma,i} \\
    \end{array}
\right],
\end{equation}
\begin{equation}
\delta \vec{u} = \vec{u}_{\rm obs} - \vec{u}_{\rm mod}.
\end{equation}
$\mu$ is the magnification matrix and $\vec{u}_{obs,i}$ is the source position counterpoint to each observed image.  

Fitting in the source plane avoids the need to solve the lens equation and allows the best model source position to be found analytically:
\begin{equation}
\vec{u}_{\rm mod} = A^{-1} \cdot {\bf b},
\end{equation}
where
\begin{equation}
A = \sum_{i}  \mu_{i}^{T} \cdot S_{i}^{-1} \cdot \mu_{i} 
\end{equation}
and 
\begin{equation}
{\bf b} = \sum_{i} \mu_{i}^{T} \cdot S_{i}^{-1} \cdot \mu_{i} \cdot \vec{u}_{{\rm obs,}i}.
\end{equation}

The image plane fitting uses a $\chi^2$ 
\begin{equation}
  \chi^2_{\rm img} = \sum_i \delta{\bf x}_i^T \cdot S_i^{-1} \cdot \delta{\bf x}_i  + \delta{\bf x}_{\rm center}^T \cdot S_i^{-1} \cdot \delta{\bf x}_{\rm center}
\end{equation}
where
\begin{equation}
\delta{\bf x}_i = {\bf x}_{{\rm obs},i}-{\bf x}_{{\rm mod},i}
\end{equation}
and 
\begin{equation}
\delta{\bf x}_{{\rm center},i} = {\bf x}_{\rm center,obs}-{\bf x}_{\rm center,mod}
\end{equation}
and the sum extends over all images, ${\bf x}_{{\rm obs},i}$ and
${\bf x}_{{\rm mod,}i}$ are the observed and modeled positions of image $i$, and ${\bf x}_{\rm center,obs}$ and
${\bf x}_{\rm center,mod}$ are the observed and modeled positions of the center of the lens potential.
The astrometric uncertainties for image $i$ are described by the
covariance matrix.  The covariance matrix we use is
\begin{equation}
S_{i} = \left[   
\begin{array}{cc}
      \sigma^2_i& 0 \\
      0 & \sigma^2_i \\
    \end{array}
\right].
\end{equation}

\bibliography{research}

\end{document}